\documentclass[graybox]{svmult}

\usepackage[bookmarks,colorlinks]{hyperref}

\usepackage{type1cm}        
\usepackage{makeidx}         
\usepackage{graphicx}        
\usepackage{multicol}        
\usepackage[bottom]{footmisc}

\usepackage{newtxtext}       %
\usepackage[varvw]{newtxmath}       

\makeindex             

\begin{document}

\title*{RIS-Parametrized Rich-Scattering Environments: Physics-Compliant Models, Channel Estimation, and Optimization}
\titlerunning{RIS-Parametrized Rich-Scattering Environments}
\author{Philipp del Hougne\orcidID{0000-0002-4821-3924}}
\institute{Philipp del Hougne \at Univ Rennes, CNRS, IETR - UMR 6164, F-35000, Rennes, France\\ \email{philipp.del-hougne@univ-rennes.fr}}
%
%
\maketitle

\abstract*{The tunability of radio environments with reconfigurable intelligent surfaces (RISs) enables the paradigm of smart radio environments in which wireless system engineers are no longer limited to only controlling the radiated signals but can in addition also optimize the wireless channels. Many practical radio environments include complex scattering objects, especially indoor and factory settings. Multipath propagation therein creates seemingly intractable coupling effects between RIS elements, leading to the following questions: How can a RIS-parametrized rich-scattering environment be modelled in a physics-compliant manner? Can the parameters of such a model be estimated for a specific but unknown experimental environment? And how can the RIS configuration be optimized given a calibrated physics-compliant model? This chapter summarizes the current state of the art in this field, highlighting the recently unlocked potential of frugal physical-model-based open-loop control of RIS-parametrized rich-scattering radio environments.}

\abstract{The tunability of radio environments with reconfigurable 
intelligent surfaces (RISs) enables the paradigm of smart radio environments in which wireless system engineers are no longer limited to only controlling the radiated signals but can in addition also optimize the wireless channels. Many practical radio environments include complex scattering objects, especially indoor and factory settings. Multipath propagation therein creates seemingly intractable coupling effects between RIS elements, leading to the following questions: How can a RIS-parametrized rich-scattering environment be modelled in a physics-compliant manner? Can the parameters of such a model be estimated for a specific but unknown experimental environment? And how can the RIS configuration be optimized given a calibrated physics-compliant model? This chapter summarizes the current state of the art in this field, highlighting the recently unlocked potential of frugal physical-model-based open-loop control of RIS-parametrized rich-scattering radio environments.}

\section{Introduction}

The ability to deterministically tune wireless channels with reconfigurable metasurfaces, often referred to as reconfigurable intelligent surfaces (RISs), is expected to play a major role in next-generation wireless networks~\cite{subrt2012intelligent,Liaskos_Visionary_2018,del2019optimally,di2019smart}. Controlling the wireless channels in addition to the input signals is a significant paradigm shift in communications. Many established tools and wisdoms from the era in which only the input signals could be controlled cannot be straightforwardly applied within this new paradigm of ``smart radio environments''. A striking example thereof is the topic of the present chapter: the treatment of (rich) scattering within the radio environment~\cite{GeorgeMag}. Traditionally, a channel matrix with suitable statistics (Rayleigh, Rician, Nakagami, etc.) is chosen to capture the effects of scattering and fading. Within smart (i.e., RIS-parametrized) rich-scattering radio environments, the situation is considerably more complicated: the dependence of the channels on the RIS configuration is non-linear due to coupling between the RIS elements, and this coupling depends on the RIS elements' proximity \textit{and} the deterministic details of the scattering environment~\cite{rabault2023tacit}. It is generally impossible to treat the impact of the RIS and the scattering environment on the wireless channel separately (although this approach is still common in theoretical works based on so-called ``cascaded channel models'').

Why is the consideration of (rich) scattering environments, as opposed to free space, important for the field of smart radio environments? 
The amount of scattering in a radio environment depends on the latter's material composition and geometry as well as the considered frequency range. The amount of scattering is often not negligible, i.e., the radio environment can often \textit{not} be approximated as being free space. 
This applies in particular to many indoor and factory radio environments of interest, and especially within the sub-6~GHz regime. Recent trends to explore millimeter-wave and terahertz regimes are still confronted with major challenges related to signal generation and attenuation, while the sub-6~GHz regime is expected to continue to play a pivotal role as part of all-spectra-integrated next-generation networks~\cite{you2021towards}. 

The essential ingredient to understanding, characterizing and optimizing an RIS-parametrized rich-scattering radio environment is a physics-compliant end-to-end channel model. We survey polarizability-based and impedance-based formulations of such channel models in Sec.~\ref{sec_ChannelModeling}, including a discussion of the nature of the non-linearity in the mapping from RIS configuration to channel, as well as a derivation of the tacit linearity assumption made by widespread but unphysical cascaded models. Having developed an understanding of physics-compliant RIS-parametrized rich-scattering channel models in Sec.~\ref{sec_ChannelModeling}, we proceed in Sec.~\ref{sec_ChannelEstimation} with the estimation of the involved parameters for a specific but $\textit{unknown}$ experimental setting. We highlight in particular the favorable inductive bias of physics-compliant channel estimation, and that built-in physical constraints enable surprisingly frugal physics-compliant channel estimation, e.g., non-coherent channel estimation or the estimation of unseen channels. Finally, having characterized a given experimental setting in Sec.~\ref{sec_ChannelEstimation}, we discuss how to optimize the RIS configuration for a desired wireless functionality in Sec.~\ref{sec_Optimization}. We present a taxonomy of different optimization objectives in terms of the role of the RIS (channel shaping vs information encoding) and the application (wireless communications, wave-based computing, sensing). We also present a taxonomy of different algorithmic optimization strategies (iterative optimization, dictionary search, adjoint method, etc.) based on closed-loop or open-loop forward mappings from RIS configuration to channel. Moreover, we describe how to efficiently evaluate the channels for different RIS configurations in a given setting with a physics-compliant model. We close in Sec.~\ref{sec_summary} with a summary and an outlook to open questions for future research.

This chapter is based on the state of the art on RIS-parametrized rich-scattering wireless channels in the Fall of 2023. The field is still rapidly evolving. Importantly, most of the theory and algorithms covered in this chapter apply more generally to any massively parametrized complex medium (MPCM)~\cite{sol2023experimentally}. RIS-parametrized rich-scattering radio environments are currently the most prominent example of MPCMs, but emerging dynamic metasurface antennas (DMAs)~\cite{sleasman2016microwave,sleasman2020implementation,antenna_patent} and wave-based signal processors~\cite{sol2022meta,sol2023reflectionless} equally rely on MPCMs~\cite{sol2023experimentally}. In fact, MPCMs emerge yet more generally across scales and wave phenomena as new approach to controlling wave-matter interactions besides the established approaches of metamaterial engineering (i.e., designing the entire system from scratch) and wavefront shaping (i.e., designing the impinging wavefront). Some experiments were already reported in nanophotonics~\cite{bruck2016all,dinsdale2021deep,delaney2021nonvolatile}, optics~\cite{resisi2019wavefront,eliezer2023tunable,li2023adaptive} and room-acoustics~\cite{ma2018shaping,wang2022controlling}. Therefore, many of the tools discussed in this chapter may soon play a role not only in next-generation wireless networks but more generally in the broader field of MPCMs.

\textit{Notation.}  $\mathbf{I}_a$ denotes the $a \times a$ identity matrix. $\mathbf{A} = \mathrm{diag}(\mathrm{a})$ denotes that $\mathbf{A}$ is a diagonal matrix constructed from the vector $\mathbf{a}$. $\left[ \mathbf{A} \right]_\mathcal{BC}$ denotes the block of the matrix $\mathbf{A}$ selected by the sets of indices $\mathcal{B}$ and $\mathcal{C}$. $\delta_{i,j}$ denotes the Kronecker delta.

\section{Channel Modeling}
\label{sec_ChannelModeling}

Throughout this chapter, we are exclusively concerned with linear time-invariant reciprocal systems. Although the RIS enables system reconfigurability, the system is static (i.e., time-invariant) during any given measurement. Moreover, for simplicity, we neglect noise throughout this chapter.

The key quantity of interest is the end-to-end wireless channel matrix $\mathbf{H}(f) \in \mathbb{C}^{N_\mathrm{R}\times N_\mathrm{T}}$ that describes the \textit{linear} input-output relation between the input wavefront $\mathbf{x}(f) \in \mathbb{C}^{N_\mathrm{T}\times 1}$ injected via the $N_\mathrm{T}$ transmitting antennas and the output wavefront $\mathbf{y}(f) \in \mathbb{C}^{N_\mathrm{R}\times 1}$ exiting the system via the $N_\mathrm{R}$ receiving antennas:
\begin{equation}
    \mathbf{y}(f) = \mathbf{H}(f) \mathbf{x}(f),
\end{equation}
where $f$ denotes frequency. $\mathbf{H}(f)$ is one block of the system's scattering matrix $\mathbf{S} \in \mathbb{C}^{N_\mathrm{A} \times N_\mathrm{A}}$, where $N_\mathrm{A}=N_\mathrm{T}+N_\mathrm{R}$, that fully characterizes the scattering of waves within the system:
\begin{subequations}
    \begin{equation}
        \mathbf{S}(f) =  \begin{bmatrix} 
	\mathbf{R^{in}}(f)  & \mathbf{H}^T(f) \\	\mathbf{H}(f)  & \mathbf{R^{out}}(f)
 \end{bmatrix},
    \end{equation} \label{eqSpartition}
    \begin{equation}
    \mathbf{H}(f) = \left[ \mathbf{S}(f)\right]_\mathcal{RT},
    \label{relationHS}
\end{equation}
\end{subequations}
where $\mathcal{R}$ and $\mathcal{T}$ denote the sets of indices assigned to the receiving antennas and transmitting antennas, respectively. $\mathbf{R^{in}}(f)\in \mathbb{C}^{N_\mathrm{T} \times N_\mathrm{T}}$ is the reflection matrix for the transmitting antennas, i.e., $\mathbf{R^{in}}(f)\mathbf{x}(f)$ gives the wavefront reflected back into the transmission lines attached to the transmitting antennas upon injection of $\mathbf{x}(f)$. Similarly, $\mathbf{R^{out}}(f)\in \mathbb{C}^{N_\mathrm{R} \times N_\mathrm{R}}$ is the reflection matrix for the receiving antennas.

Given that our system (i.e., the radio environment) is parametrized by the RIS, the key question is now to understand how $\mathbf{H}(f)$ depends on the RIS configuration $\mathbf{c}(f)\in\mathbb{C}^{N_\mathrm{S}\times 1}$ (to be defined more carefully below), where $N_\mathrm{S}$ denotes the number of RIS elements. In other words, we seek a \textbf{forward model} $\mathcal{F}$ for the mapping from RIS configuration $\mathbf{c}(f)$ to wireless channel matrix $\mathbf{H}(f,\mathbf{c}(f))$:
\begin{equation}
\mathcal{F}: \mathbf{c}(f) \mapsto \mathbf{H}(f,\mathbf{c}(f)).
\end{equation}
As we will see in the following subsections, this mapping $\mathcal{F}$ is in general non-linear (which does \textit{not} contradict the fact that $\mathbf{H}(f,\mathbf{c}(f))$ is itself a linear mapping from $\mathbf{x}(f)$ to $\mathbf{y}(f)$). In some cases, one may consider to ``blindly'' learn a neural surrogate forward model, i.e., to approximate the mapping from $\mathbf{c}(f)$ to  $\mathbf{H}(f,\mathbf{c}(f))$ with an artificial neural network (ANN). However, our goal in this section is to formulate closed-form forward models derived from first physical principles in order to gain insights into the inner workings of this mapping. Moreover, as we show in Sec.~\ref{sec_ChannelEstimation}, physics-compliant models yield more compact and more accurate forward models than neural surrogate forward models, and their parameters can be estimated with surprisingly frugal methods.

The wireless entities of primary interest in a rich-scattering RIS-parametrized radio environment are the $N_\mathrm{A}$ antennas via which waves enter or exit the system and the $N_\mathrm{S}$ RIS elements via which  the system's transfer function can be tuned. Both antennas and RIS elements are naturally discrete. 
The challenge in formulating a physics-compliant forward model is how to capture the coupling between these $N_\mathrm{P} = N_\mathrm{A}+N_\mathrm{S} $ primary wireless entities. This coupling depends on their spatial arrangement but also on the rich-scattering radio environment. If the radio environment was free space, the coupling of the primary wireless entities would only depend on their spatial arrangement and close-by entities would generally experience stronger coupling. In contrast, the additional reverberation-induced coupling under rich-scattering conditions is of long-range nature: as waves bounce around a rich-scattering environment, they may encounter different RIS elements along their trajectory, irrespective of how close these RIS elements are to each other. Reverberation-induced coupling can be stronger than proximity-induced coupling in certain settings~\cite{rabault2023tacit}. The most important insight into the difference between free space and rich-scattering radio environments is that in the latter, the coupling between the primary wireless entities is deterministically modified by the rich scattering~\cite{PhysFad,rabault2023tacit,prod2023efficient,mursia2023modeling,sol2023experimentally,tapie2023systematic}.

Two essentially equivalent approaches to capturing these coupling effects are presented in the subsequent subsections. In Sec.~\ref{subsec_polarizability_based_model}, we describe the primary wireless entities as dipoles characterized by their polarizabilities (not to be confused with polarizations) and coupled via \textit{background} Green's functions that depend on the primary dipoles' spatial arrangement and the scattering environment~\cite{sol2023experimentally}. Polarizability is a local concept\footnote{Polarizability being a local concept means that a given dipole's polarizability value does not depend on anything happening anywhere else other than at the given dipole's location.} that we consider ideally suited to capture the local scattering properties of the system defined by the RIS configuration. Meanwhile, the Green's functions capture the non-local coupling effects which change upon any perturbation anywhere in the system~\cite{sol2023experimentally}. 
Because we perceive the polarizability-based approach as particularly transparent, compact and insightful, we develop it in more detail. In particular, we derive a closed-form expression for the background Green's functions in the special case in which the scattering environment is composed of dipoles surrounded by free space, and we derive the assumptions about the truncation of multi-bounce paths that are tacitly made by unphysical but widespread cascaded models~\cite{rabault2023tacit}.
In Sec.~\ref{subsec_impedance_based_model}, we derive an alternative impedance-based formulation that treats the RIS elements as auxiliary ports terminated with tunable load impedances. Here, the coupling effects are lumped into the ports' self and mutual coupling coefficients, accounting for proximity and scattering in the radio environment~\cite{tapie2023systematic}. An important insight, applicable to either formulation, is that there is no need for an explicit description of the scattering environment, as long as its impact on the coupling between the primary wireless entities is correctly captured~\cite{sol2023experimentally,tapie2023systematic}.

\subsection{Polarizability-Based Model}
\label{subsec_polarizability_based_model}

\subsubsection{Model Formulation}
\label{subsubsec_polarizability_model_formulation}

In the polarizability-based formulation, each primary wireless entity is modeled as a dipole characterized by its polarizability~\cite{PhysFad,prod2023efficient,sol2023experimentally}. The polarizability $\alpha_i$ of the $i$th dipole relates its induced dipole moment $p_i$ to the magnitude $E_i$ of the incident field at the $i$th dipole's location and along the $i$th dipole's orientation:
\begin{equation}
    p_i(f) = \alpha_i(f) E_i(f).
    \label{eq_pi}
\end{equation}
The field $E_i$ is the superposition of the ``external'' field due to the incoming wavefront and the fields that are re-radidated by the  dipoles:
\begin{equation}
    E_i(f) = E_i^\mathrm{ext}(f) + \sum_{i=1}^{N_\mathrm{P}} G_{ij}(f)p_j(f),
    \label{eq_Ei}
\end{equation}
where 
$ E_i^\mathrm{ext}(f)$ is the component of the external field at the location of the $i$th dipole and along the latter's orientation ($ E_i^\mathrm{ext}(f)$ is non-zero only for $i \in \mathcal{A} = \mathcal{T} \cup \mathcal{R}$) and 
$G_{ij}(f)$ is the frequency-dependent \textit{background} Green's function between the positions of the dipoles indexed $i$ and $j$~\cite{sol2023experimentally}. Specifically, $G_{ij}(f)$ is the component of the field along the $i$th dipole's orientation at the location of the $i$th dipole due to a unit dipole moment at the location of the $j$th dipole with the same orientation as the $j$th dipole. Because we assume reciprocity, $G_{ij}(f)=G_{ji}(f)$. The \textit{background} Green's function accounts for the radio environment's full complexity~\cite{sol2023experimentally}.  If our radio environment was simply free space, $G_{ij}(f)$ would simplify to the free-space Green's function for which closed-form expressions exist and $G_{ii}(f)=0$. However, since we are interested in rich-scattering radio environments, we work with the \textit{background} Green's function for which in general no closed-form expression exists. For the special case of the background medium being itself composed of discrete dipoles surrounded by free space, we derive a closed-form expression for the background Green's function in Sec.~\ref{subsec_DipoleScattEnv}. In general, $G_{ii}(f) \neq 0$ for complex radio environments because there exist paths starting at the $i$th dipole and returning to the $i$th dipole that only bounce off environmental scattering objects without encountering any of the primary wireless entities along their trajectory. Such paths give rise to so-called ``self-interactions''.

To self-consistently solve the system of coupled equations defined by Eq.~(\ref{eq_pi}) and Eq.~(\ref{eq_Ei}) for $p_i$, we solve Eq.~(\ref{eq_pi}) for $E_i(f)$, insert the result into Eq.~(\ref{eq_Ei}),
\begin{equation}
    \alpha_i^{-1}(f)p_i(f) = E_i^\mathrm{ext}(f) + \sum_{i=1}^{N_\mathrm{P}} G_{ij}(f)p_j(f),
\end{equation}
and adopt a matrix notation (dropping the frequency dependence for conciseness):
\begin{equation}
    \mathbf{A} \mathbf{p} = \mathbf{E}^\mathrm{ext} + \mathbf{G}\mathbf{p},
    \label{eqstep3}
\end{equation}
where 
$\mathbf{A} = \mathrm{diag}\left(\left[\alpha_1^{-1},\alpha_2^{-1},\dots,\alpha_{N_\mathrm{P}}^{-1}\right]\right) \in \mathbb{C}^{N_\mathrm{P} \times N_\mathrm{P}} $ is a diagonal matrix containing the dipoles' inverse polarizabilities, 
$\mathbf{p} = \left[p_1,p_2,\dots,p_{N_\mathrm{P}}\right] \in \mathbb{C}^{N_\mathrm{P} \times 1} $ is a vector containing the dipoles' dipole moments, 
$\mathbf{E}^\mathrm{ext} = \left[E^\mathrm{ext}_1,E^\mathrm{ext}_2,\dots,E^\mathrm{ext}_{N_\mathrm{P}}\right] \in \mathbb{C}^{N_\mathrm{P} \times 1} $ is a vector containing the external fields incident at the locations of the dipoles along their orientations (recall that $E^\mathrm{ext}_i = 0\ \forall \ i \in \mathcal{S}$, where $\mathcal{S}$ is the set of dipole indices associated with RIS elements),
and $\mathbf{G} \in \mathbb{C}^{N_\mathrm{P} \times N_\mathrm{P}}$ is a matrix whose $(i,j)$th entry is $G_{ij}$. $\mathbf{G}$ is symmetric (given our reciprocity assumption) and in general \textit{not} hollow. Solving Eq.~(\ref{eqstep3}) for $\mathbf{p}$, we obtain
\begin{equation}
    \mathbf{p} = \left( \mathbf{A} - \mathbf{G} \right)^{-1} \mathbf{E}^\mathrm{ext} = \mathbf{W}^{-1} \mathbf{E}^\mathrm{ext},
    \label{eq4}
\end{equation}
where we define our system's \textbf{interaction matrix} $\mathbf{W} = \mathbf{A} - \mathbf{G} \in \mathbb{C}^{N_\mathrm{P} \times N_\mathrm{P}} $. The inversion of the interaction matrix in Eq.~(\ref{eq4}) compactly captures the infinite number of multi-bounce paths, as we will see in more detail in Sec.~\ref{subsec_rabu}.

At this stage, we should clarify how $\mathbf{W}$ is related to the RIS configuration $\mathbf{c}$. We stated previously that the RIS configuration defines the local scattering properties at the locations of the RIS elements, i.e., the polarizability values of the dipoles representing RIS elements. Specifically, within the polarizability-based framework in this Sec.~\ref{subsec_polarizability_based_model}, we define $\mathbf{c}$ as the vector containing the inverse polarizabilities of the dipoles representing RIS elements:
\begin{equation}
    \mathbf{c} = \left[ \alpha_i^{-1} \ \middle| \  i \in \mathcal{S} \right]. 
    \label{def_c_polariz}
\end{equation}
We can now partition $\mathbf{W}$ into $2 \times 2$ blocks as follows:
\begin{equation}
    \mathbf{W} = \mathbf{A} - \mathbf{G} = \begin{bmatrix}  \alpha_\mathrm{A}^{-1}\mathbf{I}_{N_\mathrm{A}} & \mathbf{0}_\mathcal{AS}   \\	 \mathbf{0}_\mathcal{SA} & \mathrm{diag}(\mathbf{c})
 \end{bmatrix} - \begin{bmatrix} 	 \mathbf{G}_\mathcal{AA} & \mathbf{G}_\mathcal{AS}   \\	 \mathbf{G}_\mathcal{SA} & \mathbf{G}_\mathcal{SS}
 \end{bmatrix},
 \label{eq_W_c}
\end{equation}
where we assume for simplicity that all antennas have the same polarizability $\alpha_\mathrm{A}$ and $\mathbf{0}_\mathcal{AS} = \mathbf{0}_\mathcal{SA}^T$  denotes an $N_\mathrm{A} \times N_\mathrm{S}$ matrix whose entries are all zero. It is clear based on Eq.~(\ref{eq_W_c}) that the choice of RIS configuration $\mathbf{c}$ impacts a part of the diagonal of the interaction matrix $\mathbf{W}$.

Let us now finally relate the inverse interaction matrix $\mathbf{W}^{-1}$ to the sought-after end-to-end channel matrix $\mathbf{H}$, and thereby clarify the dependence of the latter on the RIS configuration $\mathbf{c}$. 
Under our assumption of identical antennas, the incoming wavefronts are proportional to $\left[\mathbf{E}^\mathrm{ext}\right]_\mathcal{A}$ and the outgoing wavefronts are proportional to $\left[\mathbf{p}\right]_\mathcal{A}$. Therefore, $\mathbf{S}$ must be equal to $\left[ \mathbf{W}^{-1} \right]_\mathcal{AA}$ up to some multiplicative and additive factors that do \textit{not} depend on $\mathbf{c}$. Working out these additional factors is not of importance here because they do not impact the functional dependence of $\mathbf{H}$ on $\mathbf{c}$, and because they cannot be determined unambiguously in a channel estimation problem, as discussed further in Sec.~\ref{sec_ChannelEstimation}. Therefore, we absorb these additive and multiplicative factors into the interaction matrix~\cite{sol2023experimentally}, and denote variables that have absorbed such factors with a circumflex:
\begin{subequations}
\begin{equation}
     \mathbf{S} = \left[ \hat{\mathbf{W}}^{-1} \right]_\mathcal{AA} =  \left[ \left( 
 \begin{bmatrix} 
	 \hat{\alpha}_\mathrm{A}^{-1}\mathbf{I}_\mathcal{AA} & \mathbf{0}_\mathcal{AS}   \\	 \mathbf{0}_\mathcal{SA} & \mathrm{diag}(\mathbf{\hat{c}})
 \end{bmatrix} - \begin{bmatrix} 
	 \hat{\mathbf{G}}_\mathcal{AA} & \hat{\mathbf{G}}_\mathcal{AS}   \\	 \hat{\mathbf{G}}_\mathcal{SA} & \hat{\mathbf{G}}_\mathcal{SS}
 \end{bmatrix}\right)^{-1} \right]_\mathcal{AA},
\end{equation}
\begin{equation}
\begin{split}
    \mathbf{H} = \left[ \hat{\mathbf{W}}^{-1} \right]_\mathcal{RT} =  \left[ \left( 
 \begin{bmatrix} 
	 \hat{\alpha}_\mathrm{A}^{-1}\mathbf{I}_\mathcal{AA} & \mathbf{0}_\mathcal{AS}   \\	 \mathbf{0}_\mathcal{SA} & \mathrm{diag}(\mathbf{\hat{c}})
 \end{bmatrix} - \begin{bmatrix} 
	 \hat{\mathbf{G}}_\mathcal{AA} & \hat{\mathbf{G}}_\mathcal{AS}   \\	 \hat{\mathbf{G}}_\mathcal{SA} & \hat{\mathbf{G}}_\mathcal{SS}
 \end{bmatrix}\right)^{-1} \right]_\mathcal{RT} \\ \propto \left[ {\mathbf{W}}^{-1} \right]_\mathcal{RT} =  \left[ \left( 
 \begin{bmatrix} 
	 {\alpha}_\mathrm{A}^{-1}\mathbf{I}_\mathcal{AA} & \mathbf{0}_\mathcal{AS}   \\	 \mathbf{0}_\mathcal{SA} & \mathrm{diag}(\mathbf{{c}})
 \end{bmatrix} - \begin{bmatrix} 
	 {\mathbf{G}}_\mathcal{AA} & {\mathbf{G}}_\mathcal{AS}   \\	 {\mathbf{G}}_\mathcal{SA} & {\mathbf{G}}_\mathcal{SS}
 \end{bmatrix}\right)^{-1} \right]_\mathcal{RT}
 \end{split},
 \label{eq_H_cb}
\end{equation}
\label{eq_H_c}
\end{subequations}

\noindent where we use the symbol $\propto$ to denote proportionality (ignoring multiplicative or additive terms, i.e., $y(x)=ax+b \propto x$). It is apparent in Eq.~(\ref{eq_H_c}) that the dependence of $\mathbf{H}$ on $\mathbf{c}$ is in general non-linear due to the matrix inversion. Therefore, widespread cascaded models assuming a linear dependence of $\mathbf{H}$ on $\mathbf{c}$ cannot be physically consistent in general, as we elaborate further in Sec.~\ref{subsec_rabu}.

The presented polarizability-based formalism applies to any arbitrary 3D setting as long as the latter is a linear time-invariant reciprocal wave system. By defining $G_{ij}$ relative to the orientation of the dipoles indexed $i$ and $j$, we have selected the relevant components from the dyadic Green's function. Our model can also be applied to situations involving antennas or RIS elements that are not well-described as a dipole by using a multi-pole expansion~\cite{lemaire1997coupled,petschulat2008multipole,alaee2018electromagnetic,majorel2022generalizing} or a collection of dipoles~\cite{bertrand2020global} to describe them. Being derived from first physical principles, no \textit{ad hoc} corrections are necessary to account for path loss, the intertwinement of amplitude and phase response of the RIS elements, frequency selectivity, or any other physical phenomenon. Frequency selectivity (dispersion) is automatically accounted for, as seen by the explicit frequency dependence that we printed at the beginning of this Sec.~\ref{subsubsec_polarizability_model_formulation}. This furthermore implies the ability to work with our model in the time domain, simply by performing an inverse Fourier transform of $\mathbf{H}(f)$~\cite{PhysFad,hugo_eucap2024}. Moreover, the intertwinement of phase and amplitude response of the RIS element is encoded in the dispersion of its polarizability, which is usually a Lorentzian function.
Some efforts recently tried to account \textit{ad hoc} for this intertwinement in otherwise unphysical models often referred to as ``practical phase shift models''~\cite{abeywickrama2020intelligent,zhang2021performance,xu2022star}. Of course, capturing this intertwinement is not enough to have a physics-compliant model because other important effects like coupling due to proximity and reverberation remain unaccounted for in such approaches. 

To summarize, a polarizability-based model describes the primary wireless entities (antennas and RIS elements) as dipoles characterized by their polarizabilities (which are tunable in the case of RIS elements) and coupled via the \textit{background} Green's functions (which account for all coupling effects, including those arising due to rich scattering). The diagonal and off-diagonal entries of the interaction matrix are the inverse polarizabilities and negatives of the background Green's functions, respectively, such that the RIS configuration appears along parts of the diagonal of the interaction matrix. The end-to-end channel matrix is proportional to a block of the inverse of this interaction matrix.

\subsubsection{Case of Environment Composed of Dipoles Surrounded by Free Space}
\label{subsec_DipoleScattEnv}

For rich-scattering radio environments, closed-form expressions for the background Green's function do not exist in general. However, for the special case of the background medium being itself composed of dipoles surrounded by free space, we derive in this subsection a closed-form expression for the background Green's functions that only depends on the polarizabilities of the environmental dipoles and the locations of all dipoles (antennas, RIS elements, environmental dipoles)~\cite{PhysFad,prod2023efficient}.

Let us assume that the scattering environment consists of $N_\mathrm{E}$ dipoles. Following Sec.~\ref{subsubsec_polarizability_model_formulation}, we can describe this setting with $N_\mathrm{P}$ primary dipoles coupled via \textit{background} Green's functions that depend on the locations and polarizabilities of the $N_\mathrm{E}$ environmental dipoles. An equivalent alternative description consists in describing the system with $N = N_\mathrm{P}+N_\mathrm{E}$ dipoles coupled via \textit{free-space} Green's functions for which closed-form expressions exist~\cite{PhysFad,prod2023efficient}. The alternative description will yield an augmented interaction matrix $\Tilde{\mathbf{W}} \in \mathbb{C}^{N \times N}$ that we can partition into a $2\times 2$ block matrix as follows:
\begin{equation}
    \mathbf{\tilde{W}} = \begin{bmatrix} 
	\mathbf{\tilde{W}}_{\mathcal{PP}}  & \mathbf{\tilde{W}}_{\mathcal{PE}} \\
	\mathbf{\tilde{W}}_{\mathcal{EP}}  & \mathbf{\tilde{W}}_{\mathcal{EE}} \\
\end{bmatrix} = \begin{bmatrix} 
	\mathbf{A}  & \mathbf{0}_{\mathcal{PE}} \\
	\mathbf{0}_{\mathcal{EP}}  & \mathbf{A}_{\mathcal{EE}} \\
\end{bmatrix} - \begin{bmatrix} 
	\mathbf{\tilde{G}}_{\mathcal{PP}}  & \mathbf{\tilde{G}}_{\mathcal{PE}} \\
	\mathbf{\tilde{G}}_{\mathcal{EP}}  & \mathbf{\tilde{G}}_{\mathcal{EE}} \\
\end{bmatrix} = \mathbf{\tilde{A}} - \mathbf{\tilde{G}},
\label{eq11}
\end{equation}
where $\mathcal{P} = \mathcal{A} \cup \mathcal{S} $, $\mathcal{E}$ denotes the set of dipole indices corresponding to the environmental dipoles, $\mathbf{A}_{\mathcal{EE}}\in\mathbb{C}^{N_\mathrm{E} \times N_\mathrm{E}}$ is a diagonal matrix containing the inverse polarizabilities of the environmental dipoles, and $\mathbf{\tilde{G}}\in \mathbb{C}^{N \times N}$ is a hollow matrix whose $(i,j)$th entry is the free-space Green's function $\tilde{G}_{ij}$. 

We can now leverage the fact that our two descriptions of the system must be equivalent, i.e.,
\begin{equation}
    \mathbf{H} \propto \left[\mathbf{W}^{-1}\right]_\mathcal{RT} = \left[\mathbf{\tilde{W}}^{-1}\right]_\mathcal{RT}
\end{equation}
or, looking at all primary dipoles, 
\begin{equation}
    \mathbf{W}^{-1} = \left[\mathbf{\tilde{W}}^{-1}\right]_\mathcal{PP}.
    \label{eq_equiv}
\end{equation}
By applying the block matrix inversion lemma to Eq.~(\ref{eq11}), we obtain~\cite{prod2023efficient}
\begin{equation}
     \left[\mathbf{\tilde{W}}^{-1}\right]_\mathcal{PP} = \left( \mathbf{A} -   \mathbf{\tilde{G}}_{\mathcal{PP}} - \mathbf{\tilde{G}}_{\mathcal{PE}} \left(\mathbf{A}_{\mathcal{EE}} -  \mathbf{\tilde{G}}_{\mathcal{EE}}\right)^{-1}\mathbf{\tilde{G}}_{\mathcal{EP}} \right)^{-1}.
     \label{eq12}
\end{equation}
Based on Eq.~(\ref{eq_equiv}), by comparing Eq.~(\ref{eq12}) with $\mathbf{W}^{-1} = \left( \mathbf{A} - \mathbf{G} \right) ^{-1} $, we find that\footnote{Isospectral reductions similar to the one underlying Eq.~(\ref{eq_closedformG}) are also explored in graph theory~\cite{bunimovich2014isospectral}, in tight-binding network engineering~\cite{longhi2016non}, and for the conception of metamaterials with hidden symmetries to enable covert scattering control~\cite{HiddenSymmetry}.}
\begin{equation}
        \mathbf{G} = \mathbf{\tilde{G}}_{\mathcal{PP}} + \mathbf{\tilde{G}}_{\mathcal{PE}} \left(\mathbf{A}_{\mathcal{EE}} -  \mathbf{\tilde{G}}_{\mathcal{EE}}\right)^{-1}\mathbf{\tilde{G}}_{\mathcal{EP}}.
        \label{eq_closedformG}
\end{equation}
Thereby, we have identified a closed-form expression for $ \mathbf{G} $ in terms of the polarizabilities of the environmental dipoles (captured by $\mathbf{A}_{\mathcal{EE}}$) and the relative locations of all dipoles (the closed-form expression of the free-space Green's functions contained in $\tilde{\mathbf{G}}$ depends on the relative distances between the dipoles)~\cite{prod2023efficient}.

Multiple insights can be derived from Eq.~(\ref{eq_closedformG}). The effect of the scattering environment on the coupling between the primary meta-atoms is captured by the second term on the right hand side of Eq.~(\ref{eq_closedformG}) that would not be there if the background medium was free space. This second term is in general a fully populated matrix, meaning that switching from free space to a scattering radio environment changes the coupling between any two primary dipoles \textit{and} causes the dipoles to self-interact~\cite{prod2023efficient,sol2023experimentally}. In free space, there are no self-interactions, implying that $\mathbf{\tilde{G}}_{\mathcal{PP}}$ is hollow. Rich scattering does enable self-interactions because the second term on the right hand side in Eq.~(\ref{eq_closedformG}) is not hollow, such that $ \mathbf{G}$ is in general not hollow.

\subsubsection{Non-Linearity in the Mapping from RIS Configuration to Channel}
\label{subsec_rabu}

So far, we have derived that $\mathbf{H}$ depends in general non-linearly on $\mathbf{c}$ because of a matrix inversion. In this Sec.~\ref{subsec_rabu}, we explore the nature and quantify the strength of this non-linearity in more detail. Thereby, we also identify the assumptions under which the polarizability-based model can specialize to the widespread (but unphysical) cascaded model.

For simplicity, we consider the special case of a scattering environment composed of dipoles surrounded by free space from Sec.~\ref{subsec_DipoleScattEnv} and chose a different $2 \times 2$ partition of the augmented interaction matrix $\mathbf{\tilde{W}}$:
\begin{equation}
    \mathbf{\tilde{W}} = \begin{bmatrix} 
	\mathbf{\tilde{W}}_{\mathcal{UU}}  & \mathbf{\tilde{W}}_{\mathcal{US}} \\
	\mathbf{\tilde{W}}_{\mathcal{SU}}  & \mathbf{\tilde{W}}_{\mathcal{SS}} \\
\end{bmatrix},
\end{equation}
where $\mathcal{U} = \mathcal{T} \cup \mathcal{R} \cup \mathcal{E}$. Using the block matrix inversion lemma, we obtain
\begin{equation}
     [\mathbf{W}^{-1}]_\mathcal{UU}  = \left( \mathbf{W}_\mathcal{UU} - \mathbf{W}_{\mathcal{US}} \mathbf{W}_{\mathcal{SS}}^{-1} \mathbf{W}_{\mathcal{SU}} \right)^{-1},
     \label{eq17}
\end{equation}
keeping in mind that $\mathbf{H} \propto [\mathbf{W}^{-1}]_\mathcal{RT}$, i.e., we ultimately seek the $\mathcal{RT}$ block of $[\mathbf{W}^{-1}]_\mathcal{UU}$. We can now express the matrix inversion in Eq.~(\ref{eq17}) as an infinite power series~\cite{rabault2023tacit}:
\begin{subequations}
\begin{equation}
     [\mathbf{W}^{-1}]_\mathcal{UU}  = \mathbf{W}_\mathcal{UU}^{-1} \sum_{k=0}^\infty \left( \mathbf{W}_{\mathcal{US}} \mathbf{W}_{\mathcal{SS}}^{-1} \mathbf{W}_{\mathcal{SU}} \mathbf{W}_\mathcal{UU}^{-1} \right)^k
\end{equation}
\begin{equation}
    =  \mathbf{W}_\mathcal{UU}^{-1} + \mathbf{W}_\mathcal{UU}^{-1}  \mathbf{W}_{\mathcal{US}} \mathbf{W}_{\mathcal{SS}}^{-1} \mathbf{W}_{\mathcal{SU}} \mathbf{W}_\mathcal{UU}^{-1} +\mathcal{O}( \mathbf{W}_\mathcal{SS}^{-2}),
\label{eq18b}
\end{equation}
\end{subequations}
where $\mathcal{O}( \mathbf{W}_\mathcal{SS}^{-2})$ denotes terms involving more than one matrix product with $\mathbf{W}_\mathcal{SS}^{-1}$. It is now apparent why the inversion of the interaction matrix compactly captures the infinite number of increasingly long multi-bounce paths. The common ratio of the infinite power series, namely $\mathbf{W}_{\mathcal{US}} \mathbf{W}_{\mathcal{SS}}^{-1} \mathbf{W}_{\mathcal{SU}} \mathbf{W}_\mathcal{UU}^{-1}$, can be physically interpreted as one bounce from $\mathcal{U}$ to $\mathcal{S}$ and back to $\mathcal{U}$. The first term of the infinite series in Eq.~(\ref{eq18b}), namely $\mathbf{W}_\mathcal{UU}^{-1}$, includes zero such bounces, the second term in Eq.~(\ref{eq18b}), namely $\mathbf{W}_\mathcal{UU}^{-1}  \mathbf{W}_{\mathcal{US}} \mathbf{W}_{\mathcal{SS}}^{-1} \mathbf{W}_{\mathcal{SU}} \mathbf{W}_\mathcal{UU}^{-1}$, includes one such bounce, etc.~\cite{rabault2023tacit}.

It should be noted that the terms $\mathbf{W}_\mathcal{UU}^{-1}$ and $\mathbf{W}_\mathcal{SS}^{-1}$ include themselves infinitely many increasingly long multi-bounce paths constrained to trajectories within $\mathcal{U}$ and $\mathcal{S}$, respectively. Let us work this out specifically for $\mathbf{W}_\mathcal{SS}^{-1}$~\cite{rabault2023tacit}:
\begin{equation}
\begin{split}
      \mathbf{W}_{\mathcal{SS}}^{-1} = \left(\mathbf{\Phi}^{-1} +  \mathbf{\mathcal{M}}_{\mathcal{SS}}\right)^{-1} = \left( \mathbf{I}_{{N}_\mathrm{S}} +  \mathbf{\Phi}\mathbf{\mathcal{M}}_{\mathcal{SS}}\right)^{-1}   \mathbf{\Phi}\\
    = \left(\sum_{k=0}^\infty 
\left( -\mathbf{\Phi}\mathbf{\mathcal{M}}_{\mathcal{SS}} \right)^{k}\right) \mathbf{\Phi} = \mathbf{\Phi} -  \mathbf{\Phi}\mathcal{M}_\mathcal{SS} \mathbf{\Phi} + \mathcal{O}\left( \mathbf{\Phi}^{2} \right) ,
\end{split}
\label{eq19}
\end{equation}
where $\mathbf{\Phi}^{-1} = \mathrm{diag}(\mathbf{c})$ captures the local scattering properties of the RIS elements (i.e., their inverse polarizabilities) and  $\mathbf{\mathcal{M}}_{\mathcal{SS}} =  \mathbf{W}_{\mathcal{SS}} - \mathbf{\Phi}^{-1}$ captures the non-local scattering properties of the RIS (i.e., the Green's functions between the RIS elements).
It is apparent that the first term of the infinite sum in Eq.~(\ref{eq19}), namely $\mathbf{\Phi}$, is diagonal and contains no interactions between RIS elements whereas the higher order terms contain increasingly strong interactions between the RIS elements.

Overall, multiple infinite series are hence involved if we unpack the dependence of $\mathbf{H}$ on $\mathbf{c}$ in terms of the underlying multi-bounce paths. This development is very insightful to explicitly point out the tacit assumptions made by linear\footnote{To be precise, Eq.~(\ref{eq_casc}) is an ``affine'' rather than a ``linear'' relation because of the constant term $\mathbf{H_0}$ but for the sake of simplicity we use the term ``linear'' throughout this chapter.} cascaded models which postulate that~\cite{huang2019reconfigurable}
\begin{equation}
    \mathbf{H}(\mathbf{c}) = \mathbf{H_0} + \mathbf{H_1} \mathbf{\Phi} \mathbf{H_2}.
    \label{eq_casc}
\end{equation}
In order to arrive from the polarizability-based model to Eq.~(\ref{eq_casc}), on the one hand, the infinite series in Eq.~(\ref{eq19}) must be truncated after the first term, i.e., we must assume $\mathcal{M}_\mathcal{SS}=\mathbf{0}$ which implies that the coupling between RIS elements due to proximity is zero. On the other hand, the infinite series in Eq.~(\ref{eq18b}) must be truncated after the second term, i.e., we must assume that there are no multi-bounce paths encountering the RIS more than once~\cite{rabault2023tacit}. Obviously, these are very strong assumptions that do not hold in general. 

At this stage, it is instructive to look closer at the paths that are not affected by the RIS configuration and hence static. Oftentimes, it is incorrectly assumed that only the line-of-sight (LOS) path is a static path. However, as developed earlier, $\mathbf{H_0}$ contains itself an infinite number of static paths bouncing around within $\mathcal{U}$ (i.e., without encountering ever the RIS). Of these, only the very first term is the LOS path and under rich-scattering conditions, multi-bounce static paths are typically very significant. Moreover, and maybe surprisingly, in general $\langle  \mathbf{H}(\mathbf{c})\rangle_\mathbf{c} \neq \mathbf{H_0} $, where $\langle   \cdot \rangle_\mathbf{c}$ denotes the average over random realizations of $\mathbf{c}$~\cite{ismail_stirring}. It is easy to see that, unless the average of the accessible polarizability values for the RIS elements is zero, $\langle  \mathbf{H_1} \mathbf{\Phi} \mathbf{H_2} \rangle_\mathbf{c} \neq \mathbf{0}$. Therefore, significant portions of the wave energy travelling along paths that do encounter the RIS can in fact remain unaffected by the RIS configuration~\cite{ismail_stirring}.

Having established the tacit truncations made by the widespread cascaded model, we can develop insights into the physical parameters determining the importance of the resulting truncation errors. Regarding the truncation of Eq.~(\ref{eq19}), the scattering cross-section, number and spatial arrangement of the RIS elements matter~\cite{rabault2023tacit}. The topology of the RIS surface impacts the latter factor, e.g., the coupling between RIS elements of a \textit{conformal} RIS is in general different from that in an otherwise identical \textit{planar} RIS. Regarding the truncation of Eq.~(\ref{eq18b}), the reverberation time of the radio environment and the dominance of the RIS therein matter~\cite{rabault2023tacit}. The longer the reverberation time is, the more significant are paths that encounter the RIS more than once. The larger the scattering cross-section and number of RIS elements are, the more dominant the RIS will be in the radio environment, meaning that more paths will encounter it more than once. The relative locations of the wireless entities also matter but general statements about this aspect are difficult to make.

Before closing this Sec.~\ref{subsec_rabu}, we introduce a metric that we find convenient to assess the amount of non-linearity in the mapping from $\mathbf{c}$ to $\mathbf{H}$, or, in other words, the truncation error made by the cascaded model. To this end, we apply multiple linear regression to the measured data in order to identify the best possible linear model; the accuracy of the latter is an upper bound to the accuracy that could be achieved with the linear cascaded model from Eq.~(\ref{eq_casc}). For a given wireless channel $H_{ij}(\mathbf{c})$, and the corresponding prediction based on the linear model $\mathring{H}_{ij}(\mathbf{c})$, we evaluate
\begin{equation}
    \zeta_{ij} = \frac{\mathrm{SD}_{\mathbf{c}}\left[H_{ij}(\mathbf{c})\right]}{\mathrm{SD}_{\mathbf{c}}\left[H_{ij}(\mathbf{c})-\mathring{H}_{ij}(\mathbf{c})\right]},
    \label{eq_zeta_definition}
\end{equation}
where $\mathrm{SD}_{\mathbf{c}}\left[\cdot\right]$ denotes the standard deviation across random realizations of $\mathbf{c}$. $\zeta_{ij}$ is hence defined analogous to a signal-to-noise ratio, where the model error plays the role of the noise. The definition in Eq.~(\ref{eq_zeta_definition}) ensures that $ \zeta_{ij}$ is independent of the static paths not affected by the RIS configuration; this is important because otherwise the value of $ \zeta_{ij}$ would be strongly influenced by the static paths that are trivial to predict whereas we are interested in the dynamic paths that are affected by the RIS, in particular those paths affected by more than one RIS element~\cite{rabault2023tacit}. The value of $\zeta_{ij}$ has a lower bound of 0~dB because trivially defining $\mathring{H}_{ij}(\mathbf{c}) = \langle  \mathbf{H}(\mathbf{c})\rangle_\mathbf{c}$, where $\langle   \cdot \rangle_\mathbf{c}$ denotes the average over random realizations of $\mathbf{c}$, achieves by definition $\zeta_{ij}=1=0\ \mathrm{dB}$.

\subsection{Impedance-Based Model}
\label{subsec_impedance_based_model}

In Sec.~\ref{subsec_polarizability_based_model} we developed a polarizability-based physics-compliant end-to-end model for RIS-parametrized rich-scattering channels. In this Sec.~\ref{subsec_impedance_based_model}, we describe an equivalent alternative model formulated in terms of impedances. Besides the scattering matrix $\mathbf{S}(f)$, an equivalent alternative description of our $N_\mathrm{A}$-port system is the impedance matrix $\mathbf{Z}(f) \in \mathbb{C}^{N_\mathrm{A} \times N_\mathrm{A}}$ that is related to $\mathbf{S}(f)$ as follows~\cite{pozar2011microwave}:
\begin{equation}
    \mathbf{S}(f) = \left( \mathbf{Z}(f) + Z_0 \mathbf{I}_{N_\mathrm{A}}\right)^{-1} \left( \mathbf{Z}(f) - Z_0 \mathbf{I}_{N_\mathrm{A}}\right),
    \label{relationSZ}
\end{equation}
where $Z_0$ is the characteristic impedance of the transmission lines attached to the antennas. To obtain $\mathbf{H}(f)$ from $\mathbf{Z}(f)$, one must first convert $\mathbf{Z}(f)$ to $\mathbf{S}(f)$ with Eq.~(\ref{relationSZ}) and then select the $\mathcal{RT}$ block of $\mathbf{S}(f)$ as in Eq.~(\ref{relationHS}). In order to derive the dependence of $\mathbf{Z}(f)$ on the RIS configuration, the RIS elements are treated as auxiliary ports that are terminated by tunable load impedances~\cite{gradoni_EndtoEnd_2020,shen2021modeling,zhang2022macromodeling,badheka2023accurate,akrout2023physically,pdh_eucap2024,tapie2023systematic}. Within the impedance-based formulation, the RIS configuration $\mathbf{c_I} \in \mathbb{C}^{N_\mathrm{S} \times 1}$ contains the load impedances terminating the auxiliary ports:
\begin{equation}
    \mathbf{c_I}(f) = \left[ \eta_i(f) \ \middle| \  i \in \mathcal{S} \right], 
    \label{def_c_impedance}
\end{equation}
where $\eta_i(f)$ is the load impedance terminating the $i$th port. Recall that only auxiliary ports are terminated by load impedances, whereas the antenna ports are connected to transmission lines through which waves enter and/or exit the system. The impedance matrix $\overline{\mathbf{Z}} \in \mathbb{C}^{N_\mathrm{P} \times N_\mathrm{P}}$ for the system including the auxiliary ports is related to $\mathbf{Z}$ as follows~\cite{tapie2023systematic}:
\begin{equation}
    \mathbf{Z}(f) = \overline{\mathbf{Z}}_\mathcal{AA}(f) - \overline{\mathbf{Z}}_\mathcal{AS}(f) \left( \overline{\mathbf{Z}}_\mathcal{SS}(f) + \mathbf{\Phi_I}(f)\right)^{-1} \overline{\mathbf{Z}}_\mathcal{SA}(f),
    \label{Zaug}
\end{equation}
where $\mathbf{\Phi_I}(f)=\mathrm{diag} \left( \mathbf{c_I}(f) \right)$. To summarize, inserting Eq.~(\ref{Zaug}) into Eq.~(\ref{relationSZ}) and selecting the $\mathcal{RT}$ block yields~\cite{tapie2023systematic}
\begin{equation}
    \mathbf{H}(\mathbf{c_i},f) = \left[ \frac{\overline{\mathbf{Z}}_\mathcal{AA}(f) - \overline{\mathbf{Z}}_\mathcal{AS}(f) \left( \overline{\mathbf{Z}}_\mathcal{SS}(f) + \mathbf{\Phi_I}(f)\right)^{-1} \overline{\mathbf{Z}}_\mathcal{SA}(f) - Z_0 \mathbf{I}_{N_\mathrm{A}}}{\overline{\mathbf{Z}}_\mathcal{AA}(f) - \overline{\mathbf{Z}}_\mathcal{AS}(f) \left( \overline{\mathbf{Z}}_\mathcal{SS}(f) + \mathbf{\Phi_I}(f)\right)^{-1} \overline{\mathbf{Z}}_\mathcal{SA}(f) + Z_0 \mathbf{I}_{N_\mathrm{A}}} \right]_\mathcal{RT},
    \label{eq_H_c_IMPEDANCE}
\end{equation}
where we use fraction notation since in a reciprocal system the order of multiplication of nominator and denominator does not matter.

The above formulation applies equally to simple free-space settings and rich-scattering radio environments~\cite{tapie2023systematic}. Changing from free space to a rich-scattering environment will change all entries of $\overline{\mathbf{Z}}$ but the equations stated above remain valid~\cite{tapie2023systematic}. Similarly to the polarizability-based formulation, there is no need to explicitly describe the scattering within the environment as long as its impact on the entries of  $\overline{\mathbf{Z}}$ is correctly captured. Analogous to Sec.~\ref{subsec_DipoleScattEnv}, closed-form expressions for $\overline{\mathbf{Z}}$ can be worked out in the special case of the scattering environment being composed of dipoles surrounded by free space~\cite{mursia2023modeling}. However, if the physics-compliant formulation is only used to describe proximity-induced coupling between the RIS elements while environmental scattering is implemented in a cascaded manner, the overall system model is \textit{not} physics-compliant because by construction it cannot capture the reverberation-induced coupling arising due to multi-bounce paths that encounter multiple RIS elements~\cite{shen2021modeling,li2022beyond,nerini2023discrete}.

\subsection{Comparison}

Compared to Eq.~(\ref{eq_H_c}) from the polarizability-based formulation, Eq.~(\ref{eq_H_c_IMPEDANCE}) from the impedance-based formulation can be perceived as being more cumbersome and less transparent in terms of understanding the dependence of the wireless end-to-end channel matrix on the RIS configuration~\cite{pdh_eucap2024}. The RIS configuration defines local scattering properties within the system. While polarizability is a local concept, impedance is a non-local concept: any perturbation of the system, no matter where, impacts all entries of the impedance matrix. Many papers related to the impedance-based formulation make simplifying assumptions (e.g., the ``unilateral approximation''~\cite{ivrlavc2010toward}) to ease the complexity of the mathematical expressions.

The decision about using the polarizability-based or the impedance-based formulation is sometimes a matter of personal choice, but in certain cases specific choices seem preferable. For deriving physical insight in terms of multi-bounce paths as we did in Sec.~\ref{subsec_rabu}, the polarizability-based formulation appears preferable~\cite{rabault2023tacit}. For experimental channel estimation problems (Sec.~\ref{sec_ChannelEstimation}) and optimizations based on experimentally estimated channels (Sec.~\ref{sec_Optimization}), the more compact polarizability-based formulation appears preferable, too~\cite{sol2023experimentally}. However, for work based on full-wave simulations, the impedance-based formulation is easier to deploy because the entries of $\overline{\mathbf{Z}}$ can be directly obtained with a single full-wave simulation, irrespective of the complexity of the radio environment~\cite{tapie2023systematic}.

\begin{important}{Key Take-Home Messages of Sec.~\ref{sec_ChannelModeling}}
\begin{enumerate}
    \item The wireless channel matrix $\mathbf{H}$ is a \textit{linear} input-output relation between $\mathbf{x}$ and $\mathbf{y}$ that depends \textit{non-linearly} on the RIS configuration $\mathbf{c}$.
    \item Coupling between any two primary wireless entities (antennas, RIS elements) arises due to their proximity as well as due to reverberation in the scattering environment. 
    \item Reverberation-induced coupling is of long-range nature: waves bouncing around in the radio environment can encounter different RIS elements along their trajectory, irrespective of how spatially close these RIS elements are.
    \item Polarizability-based physics-compliant models describe each primary wireless entity as a dipole characterized by its polarizability (which is tunable for RIS elements) and coupled to the other dipoles via \textit{background} Green's functions (which capture the effect of the scattering environment).
    \item The system's interaction matrix $\mathbf{W}$ is the sum of a diagonal matrix capturing the inverse polarizabilities and a symmetric matrix capturing the background Green's functions. The inversion of $\mathbf{W}$ compactly captures the infinite number of increasingly long multi-bounce paths. 
    \item The wireless channel matrix is proportional to one block of the inverse interaction matrix: $\mathbf{H} \propto \left[ \mathbf{W}^{-1} \right]_\mathcal{RT}$. 
    \item Alternative physics-compliant impedance-based models describe RIS elements via auxiliary ports terminated with tunable load impedances. The self-impedances and mutual-impedances capture the effect of the scattering environment.
\end{enumerate}
\end{important}

\section{Channel Estimation}
\label{sec_ChannelEstimation}

Having established physics-compliant models for RIS-parametrized rich-scattering radio environments in Sec.~\ref{sec_ChannelModeling}, we were able to gain some physical insights into the functional dependence of the wireless channel matrix $\mathbf{H}$ on the RIS configuration $\mathbf{c}$. But for any algorithmic developments based on such models to be of practical value, we must be able to estimate the parameters of these models such that they describe a given experimental situation. In other words, we need to estimate the parameters of our physics-compliant model for a given and a priori unknown radio environment. This leads to the problem of \textit{physics-compliant} end-to-end channel estimation that is to date (at the time of writing in Fall 2023) a largely uncharted area except for very recent results from Summer 2023~\cite{sol2023experimentally} that we summarize in this Sec.~\ref{sec_ChannelEstimation}. We expect significant further developments in this area in the near future.

Channel estimation for RIS-parametrized channels has previously received some theoretical attention~\cite{wang2020channel,hu2021two,alexandropoulos2021hybrid}, however, only for free-space radio environments and only based on linear cascaded models rather than physics-compliant channel models. We do not discuss such works in detail here given our focus on rich-scattering settings, and our desire to experimentally validate the algorithms which requires the use of physics-compliant models.

Successful physics-compliant end-to-end channel estimation in a given radio environment enables \textit{open-loop} wave control because the channel matrix corresponding to any conceivable RIS configuration can be accurately predicted~\cite{sol2023experimentally}. Then, the RIS configuration can be optimized for a desired wireless functionality without any additional measurements. By contrast, in the absence of a calibrated forward model, an algorithm would have to iteratively adjust the RIS configuration based on feedback measured experimentally in situ after each iteration to assess the extent to which the current channel matrix is suitable for the sought-after wireless functionality. Moreover, if a different wireless functionality was desired, one would have to run again through a similar closed-loop optimization involving additional measurements. Optimization is discussed in more detail in the dedicated Sec.~\ref{sec_Optimization}.

End-to-end channel estimation refers to approximating the forward function $\mathcal{F}$ that maps $\mathbf{c}$ to $\mathbf{H}$, i.e., $\mathcal{F}: \mathbf{c} \mapsto \mathbf{H}(\mathbf{c})$, for a specific experimental setting.
In principle, this goal can be achieved without a physics-compliant model. 
Given the overwhelming complexity of unknown rich-scattering radio environments, nowadays a tempting approach is to ``blindly'' learn a surrogate neural forward model (also referred to as ``digital twin'') by training an ANN to approximate $\mathcal{F}$~\cite{stylianopoulos2022deep,momeni2023backpropagation}. However, as we will see in Sec.~\ref{subsec_favinductbias}, physics-compliant channel estimation yields orders of magnitude more accurate and more compact forward models than such neural approaches~\cite{sol2023experimentally}. This can be traced back to the favorable inductive bias of the physics-compliant model. More importantly, we show in Sec.~\ref{subsec_frugal} that surprisingly frugal methods can successfully estimate the parameters of a physics-compliant model without ever having measured phase or without ever having measured some of the channels of interest~\cite{sol2023experimentally}. These frugal channel estimation techniques are unique to physics-compliant channel estimation. Neither a neural surrogate forward model nor closed-loop iterative schemes could optimize RIS configurations for coherent wave control to achieve desired wireless functionalities under such frugal constraints. We focus on \textit{physics-compliant} channel estimation in unknown rich-scattering conditions in this Sec.~\ref{sec_ChannelEstimation}, and only briefly refer to other approaches for benchmarking where possible.

\subsection{Principle}
\label{subsec_princple_ChanEst}

As developed in Sec.~\ref{subsubsec_polarizability_model_formulation}, irrespective of the complexity of the unknown rich-scattering radio environment, there is a compact physics-compliant closed-form expression for $\mathcal{F}$~\cite{sol2023experimentally}:

\begin{equation}
     \mathbf{H}(\mathbf{\hat{c}}) = \left[ \hat{\mathbf{W}}^{-1}(\mathbf{\hat{c}}) \right]_\mathcal{RT} =  \left[ \left( 
 \begin{bmatrix} 
	 \hat{\alpha}_\mathrm{A}^{-1}\mathbf{I}_\mathcal{AA} & \mathbf{0}_\mathcal{AS}   \\	 \mathbf{0}_\mathcal{SA} & \mathrm{diag}(\mathbf{\hat{c}})
 \end{bmatrix} - \begin{bmatrix} 
	 \hat{\mathbf{G}}_\mathcal{AA} & \hat{\mathbf{G}}_\mathcal{AS}   \\	 \hat{\mathbf{G}}_\mathcal{SA} & \hat{\mathbf{G}}_\mathcal{SS}
 \end{bmatrix}\right)^{-1} \right]_\mathcal{RT} ,
 \label{eq_27}
\end{equation}
which is reproduced from Eq.~(\ref{eq_H_cb}). All we know is that all antennas are nominally identical (i.e., they all have the same polarizability), all RIS elements are nominally identical (i.e., the same $2^q$ polarizability values are accessible for each RIS element), and the system is reciprocal. The radio environment's geometry and material composition are unknown. For concreteness, we consider in the following 1-bit programmable RIS elements (i.e., $q=1$) but the approach can straightforwardly be applied to multi-bit tunable RIS elements with $q>1$, too.

How many parameters do we need to estimate? There are $1+2^q$ local parameters (namely $\hat{\alpha}_\mathrm{A}$ and the $2^q$ possible values that the entries of $\mathbf{\hat{c}}$ can take) and $\frac{1}{2} N_\mathrm{P}(N_\mathrm{P}+1)$ non-local parameters (because $\hat{\mathbf{G}}=\hat{\mathbf{G}}^T$ due to reciprocity). These parameters are complex-valued, so in total we must estimate $2\left(1+2^q+\frac{1}{2} N_\mathrm{P}(N_\mathrm{P}+1) \right)$ values. The number of parameters to be estimated is hence $\mathcal{O}(N_\mathrm{P}^2)$ and, importantly, does \textit{not} dependent of the complexity of the radio environment~\cite{sol2023experimentally}. No explicit description of or knowledge about the radio environment is required. The effects of rich scattering on the coupling between the primary wireless entities is fully captured by the estimated entries of $\hat{\mathbf{G}}$~\cite{sol2023experimentally}.

In order to estimate the physical model's parameters, we make one-off calibration measurements in the unknown experimental setting of interest. Specifically, for a set of $m$ known RIS configurations, we measure the corresponding channel matrices. The approach taken in this Sec.~\ref{sec_ChannelEstimation} is to choose a known set of $m$ random RIS configurations for the calibration measurements, and to identify the parameters of the physics-compliant model via gradient descent with an error backpropagation algorithm~\cite{sol2023experimentally}. Note that there is an infinite number of valid parameter choices that would all serve equally well to map $\mathbf{c}$ to $\mathbf{H}$. There is no need to remove this ambiguity, nor is this in general possible. In fact, we embrace this ambiguity because it facilitates the convergence of our gradient descent~\cite{sol2023experimentally}.

\subsection{Favorable Inductive Bias of Physics-Compliant Model}
\label{subsec_favinductbias}

Having established the principle of physics-compliant end-to-end channel estimation in Sec.~\ref{subsec_princple_ChanEst}, we can now compare its performance in terms of (i) the achieved accuracy $\zeta_{ij}$, and (ii) the number of required calibration examples $m$ against two important benchmarks. The first benchmark is a linear model whose parameters we obtain via multiple linear regression. Its performance is an upper bound on the performance achievable with the linear cascaded model from Eq.~(\ref{eq_casc})~\cite{huang2019reconfigurable}. The second benchmark is a multilayer perceptron feedforward ANN that is widely used for ``blind'' function approximation without any a priori knowledge~\cite{stylianopoulos2022deep,momeni2023backpropagation}. All models are calibrated (``trained'') with the same data set described previously. The accuracy is evaluated analogous to Eq.~(\ref{eq_zeta_definition}) for each considered model (linear, neural, physics-compliant):
\begin{equation}
    \zeta_{ij} = \frac{\mathrm{SD}_{\mathbf{c}}\left[H_{ij}(\mathbf{c})\right]}{\mathrm{SD}_{\mathbf{c}}\left[H_{ij}(\mathbf{c})-\mathring{H}_{ij}(\mathbf{c})\right]},
    \label{eq_zeta_definition2}
\end{equation}
where $\mathring{H}_{ij}(\mathbf{c})$ is the prediction of the considered model and $H_{ij}(\mathbf{c})$ is the experimentally measured ground truth.

\begin{figure}[t]
\centering
\includegraphics[width=\linewidth]{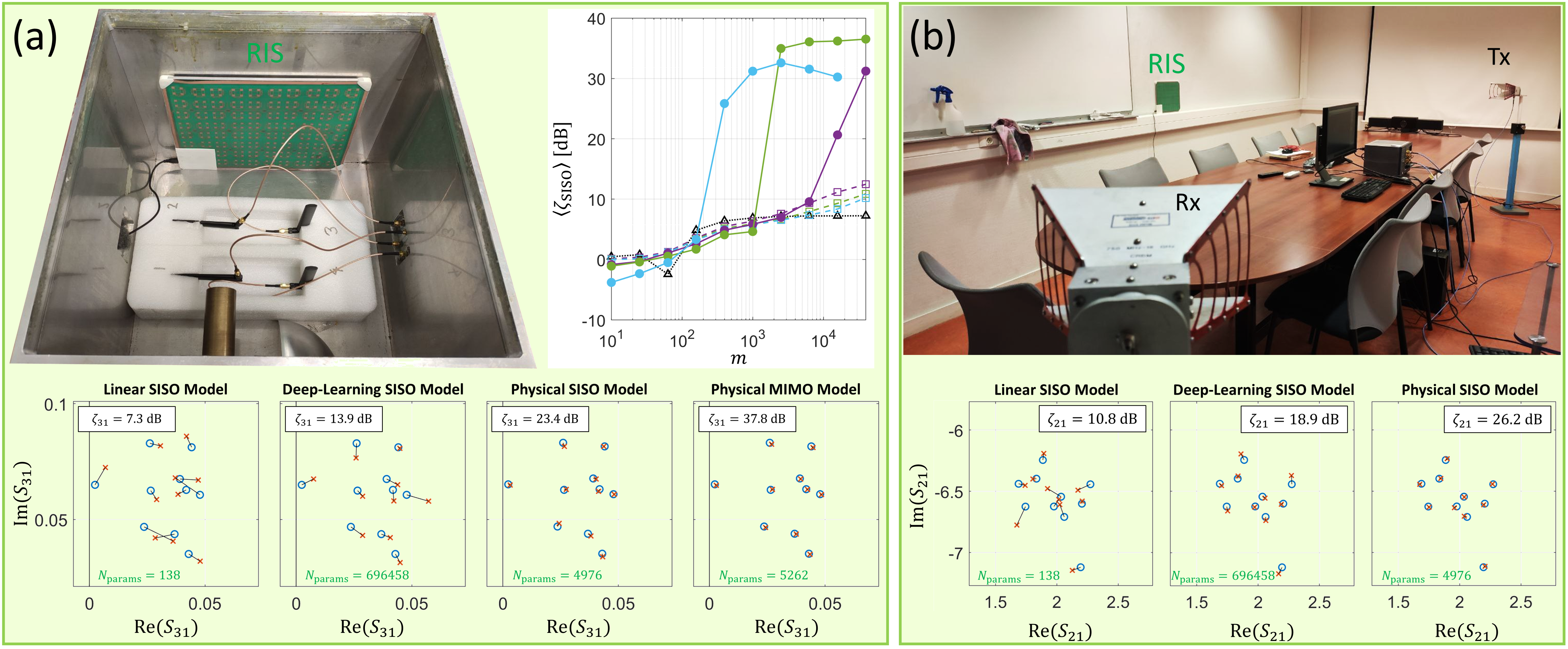}
\caption{End-to-end channel estimation in an academic rich-scattering setup (a) and a meeting room (b). The bottom row shows ground truth and predicted values of the complex-valued channel coefficients for ten unseen random RIS configurations, for different models (linear, neural, physics-compliant) calibrated with $m=4 \times 10^4$ examples. The model accuracy $\zeta_{ij}$ evaluated with Eq.~(\ref{eq_zeta_definition2}) is indicated in each case. In addition, in (a) the average achieved accuracy of the models for different amounts $m$ of calibration data are shown (triangle: linear; square: neural; circle: physics-compliant; purple: SISO; green: MIMO; cyan: entire $\mathbf{S}$). (Adapted from Ref.~\cite{sol2023experimentally}.)}
\label{fig1}
\end{figure}

Applied to the two experimental situations shown in Fig.~\ref{fig1}, an academic rich-scattering chaotic cavity and a meeting room, it is apparent that the physical model achieves at least one order of magnitude better accuracy than the neural benchmark while using two orders of magnitude fewer parameters, and it outperforms the linear benchmark by at least two orders of magnitude in terms of accuracy~\cite{sol2023experimentally}. While the poor accuracy of the linear model is not surprising in sight of the theory developed in Sec.~\ref{sec_ChannelModeling}, the inferiority of the neural model's accuracy may come as a surprise. In principle, a generic neural model can approximate any arbitrarily complex function; however, a generic neural model would require much more parameters and training data than an approach based on a valid model of the sought-after function. Moreover, it should be noted that the neural model cannot converge to the physical model because its feedforward architecture fundamentally differs from the recurrent scattering encoded in the matrix inversion of the physical model. One may speculate that this architectural difference explains the neural model's struggle to accurately map $\mathbf{c}$ to $\mathbf{H}$. Generally speaking, the more valid physics knowledge is injected into a model, the better the performance will be because the model will benefit from a more \textbf{favorable inductive bias}. Our ``pure-physics'' model in Eq.~(\ref{eq_27}) is at the extreme end of so-called ``model-based deep learning''~\cite{shlezinger2023model}, to the point that it can actually be characterized as a traditional signal processing approach without any deep learning. 

Interestingly, the dependence of the accuracy $\zeta_{ij}$ on the number of calibration examples $m$ appears to have a phase transition (see Fig.~\ref{fig1}a)~\cite{sol2023experimentally}. No rigorous theory has been worked out for this phase transition to date, but it is reminiscent of a similar phenomenon in compressed sensing~\cite{amelunxen2014living}. Strikingly, the more channels are to be predicted, the earlier this phase transition occurs~\cite{sol2023experimentally}. In other words, the more channel coefficients the physical model has to predict, the fewer calibration examples it needs. This trend is very favorable given the advent of massive multiple-input multiple-output (MIMO) wireless communications systems. Intuitively, this trend can be explained by the fact that the physical model ``understands'' the relation between different channel coefficients. By contrast, the neural model cannot ``understand'' these relations, and its performance deteriorates when it must predict more channel coefficients.

\subsection{Frugal Physics-Compliant Channel Estimation}
\label{subsec_frugal}

We have already seen in Sec.~\ref{subsec_favinductbias} that physics-compliant channel estimation drastically reduces the number of required calibration examples and model parameters while offering orders of magnitude larger accuracies. These are already significant steps toward ``frugality''. However, the naturally built-in constraints of physics-compliant models enable yet significantly more frugality.

The first type of advanced frugality that we consider is \textbf{non-coherent channel estimation}. The requirement for coherent measurements to perform channel estimation implies a significant hardware cost. Is it possible to alleviate the vexing requirement for coherent calibration measurements? In other words, is it possible to accurately estimate the physical model's parameters purely based on non-coherent (i.e., phaseless) measurements? Certainly with a neural approach this would be impossible because the ANN could not retrieve any phase relations. In the case of the physics-compliant model, however, it turns out to be possible~\cite{sol2023experimentally}. The naturally built-in constraints ``force'' the physical model to correctly predict phase relations if it correctly predicts amplitude relations. 
For the experiment from Fig.~\ref{fig1}(a), it was possible to achieve accuracy values on the order of 20~dB for all channel coefficients purely based on non-coherent measurements~\cite{sol2023experimentally}. Both the phase relations between different channel coefficients and their dependence on the RIS configuration were accurately inferred from phaseless calibration data. (Note that these accuracy values significantly exceed those achieved by the neural model  \textit{with} access to phase information, as seen in Fig.~\ref{fig1}(a).) Of course, there is a global phase constant that cannot be retrieved but this global phase constant has no physical meaning. Some care must be taken regarding the choice of pilot signals in the case of non-coherent physics-compliant channel estimation: for any given pilot, multiple transmitters should radiate energy in order to probe the phase relations between the antenna ports. In other words, the one choice of pilot signals that must be avoided are one-hot pilots that only radiated energy from one transmitter at a time. A simple suitable choice is to draw the pilot signals from a complex-valued random distribution with normally distributed real and imaginary parts. The ability to perform non-coherent channel estimation unlocks coherent wave control in unknown complex environments (in terms of both configuring the RIS \textit{and} choosing the input wavefront $\mathbf{x}$) without ever having measured phase~\cite{sol2023experimentally}.

For the second type of frugality we go one step further yet and ask if we can \textbf{estimate unseen channels}. For concreteness, we consider the task of mapping $\mathbf{c}$ to the entire scattering matrix $\mathbf{S}$ based on calibration data in which one block of $\mathbf{S}$, say $\mathbf{R^{in}}$ (see Eq.~(\ref{eqSpartition})), is excluded, i.e., never measured. In other words, some antennas only operate in receiving mode but never transmit themselves such that we have no measurements for their reflection coefficients nor for the transmission coefficients between them. For a neural model it would be impossible to make any prediction for such unseen channel coefficients. For the physics-compliant model, however, it turns out that its naturally built-in constraints are once again strong enough to ensure that all essential features of the unseen channel coefficients are correctly predicted~\cite{sol2023experimentally}. For the example from Fig.~\ref{fig1}(a), accuracies of 23~dB or better were achieved for unseen channel coefficients~\cite{sol2023experimentally} (in contrast to the neural or linear models whose accuracies remain at least one order of magnitude below this value with calibration data that \textit{does} include all coefficients). The only difficulty of the physics-compliant model regarding the prediction of unseen channel coefficients is the prediction of the non-RIS-dependent static components of the unseen channel coefficients (which does not impact our accuracy metric). This difficulty is understandable since the paths contributing to these fixed components are probed only very indirectly in the available calibration data. In some applications, this constant offset does not matter. In any case, a direct measurement for a single known RIS configuration is enough to correct the constant offset~\cite{sol2023experimentally}.

\begin{important}{Key Take-Home Messages of Sec.~\ref{sec_ChannelEstimation}}
\begin{enumerate}
\item Estimating the parameters for an end-to-end channel model to describe a specific unknown experimental setting is necessary for \textit{open-loop} wave control therein, enabling the optimization of the RIS for any desired functionality without additional measurements.
\item The number of parameters to be estimated for a physical model is independent of the complexity of the radio environment.
\item Physics-compliant channel estimation benefits from a favorable inductive bias compared to physics-agnostic neural surrogate forward models: using two orders of magnitude fewer parameters, at least one order of magnitude better accuracy is achieved in recent experiments.
\item The more channel coefficients are considered, the fewer calibration examples the physics-compliant approach requires.
\item Naturally built-in constraints of the physical model enable surprisingly frugal channel estimation methods, such as based on non-coherent measurements or without any information about some of the channels of interest. Thereby, the hardware cost of channel estimation can be lowered drastically.
\item These frugal channel estimation capabilities of physics-compliant approaches are inaccessible with neural or other physics-agnostic approaches.

\end{enumerate}
\end{important}

\section{Optimization}
\label{sec_Optimization}

In the previous two sections, we have formulated a closed-form physics-compliant model (Sec.~\ref{sec_ChannelModeling}) and estimated its parameters so that it describes a given but unknown experimental setting (Sec.~\ref{sec_ChannelEstimation}). Hence, we are now ready to optimize the RIS configuration in a given but unknown rich-scattering environment to achieve a desired wireless functionality. In this Sec.~\ref{sec_Optimization}, we provide a taxonomy of such optimizations in terms of their objectives (Sec.~\ref{subsec_tax_opt_objectives}) as well as in terms of their algorithmic strategies (Sec.~\ref{subsec_tax_opt_strateg}), and we discuss efficient algorithmic implementations based on physics-compliant models (Sec.~\ref{subsec_prod}).

\subsection{Taxonomy of Optimization Objectives}
\label{subsec_tax_opt_objectives}

Broadly speaking, there are two roles that the RIS can play. Specifically, we distinguish between \textbf{channel shaping} and \textbf{information encoding}. The difference arises with respect to how data enters and exits the system. In the case of \textbf{channel shaping}, the input data is encoded into the input wavefront $\mathbf{x}$, the output data is encoded into the output wavefront $\mathbf{y}$, and the RIS is used to shape the linear mapping $\mathbf{H}$ from $\mathbf{x}$ to $\mathbf{y}$. In this case, the mapping from input data to output data is inevitably \textbf{linear}. This is the more ``conventional'' role of the RIS. However, it is also possible to encode the input data into the RIS configuration $\mathbf{c}$ and to extract the output data from $\mathbf{H}=\mathbf{y}/\mathbf{x}$. In this case, the role of the RIS relates to \textbf{information encoding}, and the mapping from input data to output data is inevitably \textbf{non-linear} (as developed in Sec.~\ref{sec_ChannelModeling}).

\begin{figure}[h]
\centering
\includegraphics[width=\linewidth]{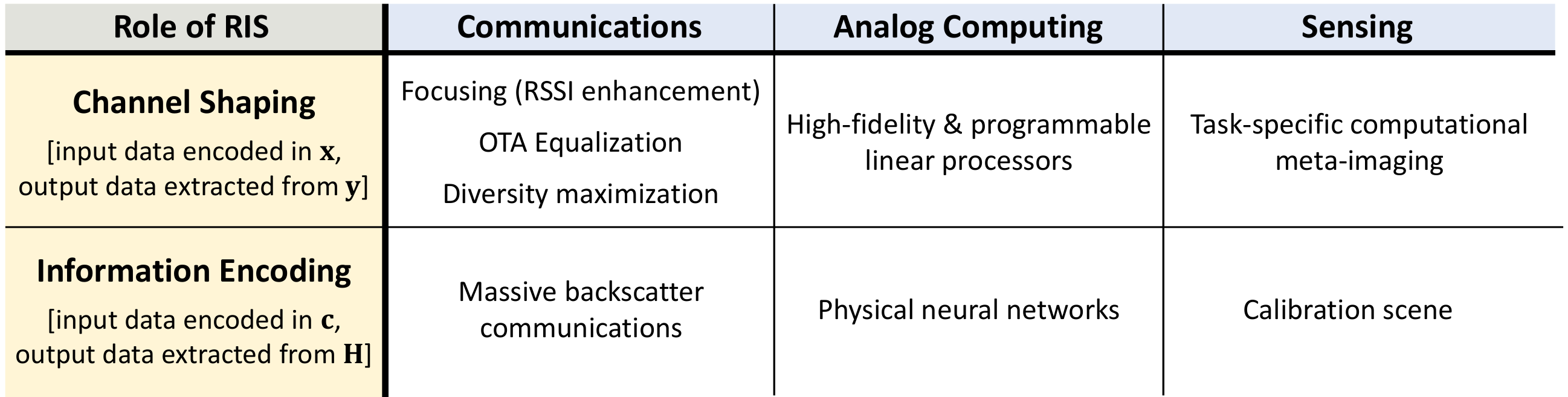}
\caption{Taxonomy of optimization objectives in RIS-parametrized rich-scattering environments.}
\label{fig2}
\end{figure}

A taxonomy of different optimization objectives is presented in Fig.~\ref{fig2}. For each of the two above-mentioned possible roles of the RIS, important applications exist, of course, in wireless communications but increasingly also in wave-based processing (computing) and sensing. Given the trend toward integrating communications, computing and sensing in future generations of wireless networks, we point out some applications to wave-based computing and sensing, too. 

Most applications explored to date fall into the realm of \textbf{wireless communications} and the RIS serves for channel shaping. First and foremost, there are many different attempts at \textit{maximizing the received signal-strength indicator (RSSI)} in rich-scattering settings by focusing waves on the receiver~\cite{Kaina_metasurfaces_2014,dupre2015wave,del2017shaping,del2016intensity}. This can be achieved by optimizing the RIS configuration to create constructive interferences at the location of the receiver. In the case of multiple transmitters, the radiated wavefront can be optimized in addition (via phase conjugation, i.e., maximum ratio transmission). Another important objective relates to \textit{over-the-air (OTA) channel equalization for resource-constrained networks} operating under rich-scattering conditions~\cite{del2016spatiotemporal,hugo_eucap2024} (e.g., for Internet-of-Things devices or Wireless Networks-on-Chip~\cite{imani2021smart,tapie2023systematic}). Here, the channel impulse response (CIR) in the time domain is considered and constructive and destructive interferences are judiciously tailored by an optimized RIS configuration in order to obtain a CIR with a single dominant tap that looks almost pulse like despite the rich scattering~\cite{hugo_eucap2024,tapie2023systematic}. Furthermore, for rich-scattering MIMO systems, the RIS configuration can be optimized in order to maximize the diversity of $\mathbf{H}$ (quantified, e.g., via the effective rank of $\mathbf{H}$)~\cite{del2019optimally,del2019optimized}. When the RIS serves for information encoding, it can serve as \textit{massive backscatter communications} device~\cite{zhao2020metasurface,f2020perfect}. Indeed, in comparison to traditional backscatter communications devices like the ``Great Seal Bug``~\cite{brooker2013lev}, RFID tags~\cite{roberts2006radio} or ambient backscatter setups~\cite{liu2013ambient}, an RIS offers orders of magnitude more degrees of freedom and a much larger aperture, enabling the implementation of advanced modulation schemes and improved security~\cite{zhao2020metasurface}. In particular, by optimizing for very special ``perfect-absorption'' conditions, physical-layer security can be achieved in RIS-based massive backscatter communications inside rich-scattering environments~\cite{f2020perfect}.

A second thread of applications of RIS in rich-scattering environments relates to analog wave-based computing~\cite{del2018leveraging}. If the RIS serves for channel shaping, it enables access to high-fidelity and in situ reprogrammable analog linear computing, which has been demonstrated in particular for signal differentiation~\cite{sol2022meta} and reflectionless routing~\cite{sol2023reflectionless}. The role of the rich-scattering here is to add significant non-local interactions between the RIS elements that boost their impact on the transfer function and hence the fidelity with which a desired transfer function can be implemented. Another way of interpreting this is that the longer the wave reverberates, the more often it revisits the RIS and hence the more sensitive it becomes to the latter's configuration~\cite{del2021deeply}. Such RIS-based wave processors implement a desired linear mathematical operation at the speed of light, and significant parallelization is possible thanks to the linearity of the wave equation~\cite{sol2022meta}. On the other hand, if the RIS serves for information encoding, the mapping from RIS to transfer function can be exploited to implement non-linear analog functions at low signal power levels, as required by energy-efficient physical neural networks (PNNs)~\cite{momeni2023backpropagation}. However, in this case, some conversions between the digital and analog domains are necessary which might limit the speed and energy efficiency.

Finally, a third thread of applications of RIS relates to sensing~\cite{saigre2022intelligent}. Sensing is broadly concerned with extracting information about a scene based on how it scatters incident waves. Thereby, it encompasses imaging, localization, detection, recognition, etc. If the RIS is used for channel shaping, it can enable the generation of mutually orthogonal~\cite{del2019optimized} or end-to-end optimized \textit{task-specific wavefronts}~\cite{del2020learned,li2020intelligent,saigre2022intelligent,qian2022noise} to probe the scene, of which in particular the latter can lead to considerable improvements in latency and other relevant metrics~\cite{del2020learned,li2020intelligent,saigre2022intelligent,qian2022noise}. In addition, the reverberation provides a ``generalized interferometric sensitivity'' that enables orders of magnitude better resolution than in free space~\cite{del2021deeply}. If the RIS is used for information encoding, it could act as programmable \textit{calibration scene} although this remains uncharted territory to date.

\subsection{Taxonomy of Algorithmic Strategies for Optimization}
\label{subsec_tax_opt_strateg}

For any of the optimization objectives outlined in Sec.~\ref{subsec_tax_opt_objectives}, a suitable cost function $\mathcal{C}(\mathbf{H}(\mathbf{c}))$ can be defined that must be minimized by optimizing the RIS configuration $\mathbf{c}$ in order to reach the desired wireless functionality as closely as possible. Oftentimes, additional constraints (e.g., regarding the RIS elements' 1-bit programmability) must be accounted for, leading to the following type of inverse design problem:
\begin{subequations}
\begin{equation}
    \mathrm{min}_\mathbf{c} \ \mathcal{C}(\mathbf{H}(\mathbf{c})) 
    \label{eqA}
\end{equation}    
\begin{equation}
    \mathrm{s.t.} \ [\mathbf{c}]_i \in \{\alpha_0^{-1},\alpha_1^{-1} \},
    \label{eqB}
\end{equation}
\end{subequations}
where $\alpha_0$ and $\alpha_1$ denote the two possible polarizability values available for dipoles representing RIS elements in the case of 1-bit programmability.

Leaving aside the constraints from Eq.~(\ref{eqB}) for a moment, one may be tempted to suspect that by inversing the forward mapping $\mathcal{F}: \mathbf{c} \mapsto \mathbf{H}(\mathbf{c})$, i.e., by formulating the inverse model $\mathcal{I}: \mathbf{H}(\mathbf{c}) \mapsto \mathbf{c}$, the inverse design problem is solved~\cite{frazier2022deep}. Unfortunately, inverse problems are generally ill-posed. On the one hand, there is no guarantee that the desired channel matrix is even physically realizable. On the other hand, if it is, there is no guarantee that there is only one RIS configuration that realizes it. These concerns of ``existance'' and ``uniqueness'' explain the common difficulty of solving the inverse problem and why there is typically no closed-form solution to it (even though we have a closed-form forward model)~\cite{wiecha2021deep,khatib2021deep}.

As an aside, sensing problems (see brief discussion in Sec.~\ref{subsec_tax_opt_objectives}) also solve inverse problems: what configuration\footnote{In the sensing context, ``configuration'' refers broadly to anything defining the structure of the scattering system: the locations and/or properties of the wireless entity.} explains the measurements? However, sensing problems differ from the inverse-design problems we discuss in this Sec.~\ref{subsec_tax_opt_strateg} in that (i) there is no ``existance'' concern since sensing problems are based on measured rather than desired channel coefficients, and (ii) a suitably diverse measurement scheme can avoid the ``uniqueness'' concerns (and dedicated research efforts seek to find suitable schemes based on spatial, spectral, or configurational diversity~\cite{del2018precise,del2019optimized,saigre2022intelligent}). Hence, while an inverse model $\mathcal{I}$ can solve sensing problems, it is usually not enough to solve inverse-design problems. 

\begin{figure}[t]
\centering
\includegraphics[width=\linewidth]{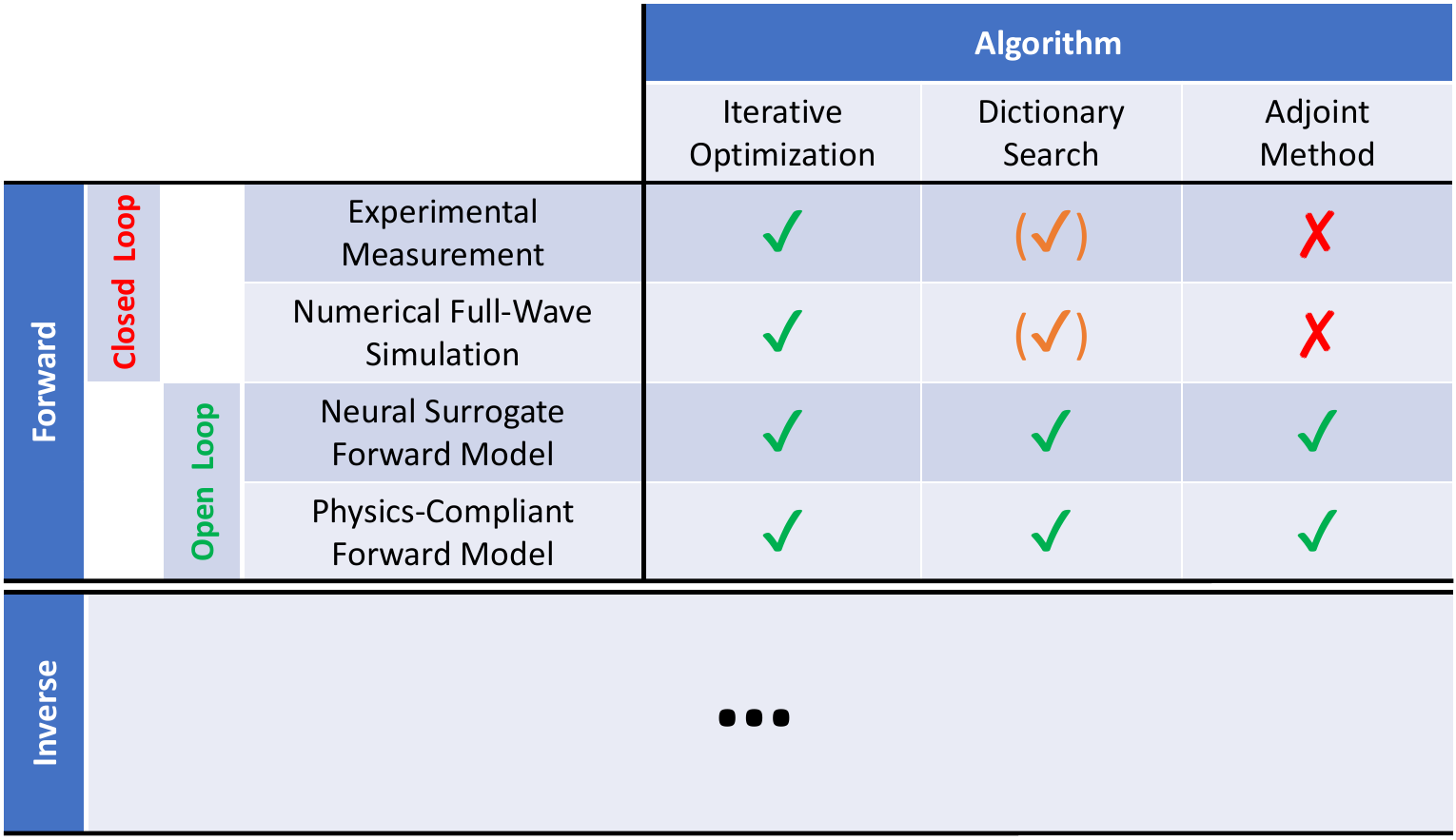}
\caption{Taxonomy of algorithmic optimization strategies in RIS-parametrized rich-scattering environments, focusing on those based on forward mappings. Note that these strategies can to some extent be combined, e.g., a dictionary search can initialize an iterative optimization or adjoint method.}
\label{fig3}
\end{figure}

A taxonomy of four major families of approaches to tackle inverse-design problems is presented in Fig.~\ref{fig3}. Within the realm of approaches using forward mappings, which is the focus of our discussion, we can distinguish between closed-loop and open-loop forward mappings. A direct experimental measurement of the channel matrix corresponding to a RIS configuration of interest is the most important example thereof. Numerical full-wave simulations are another but rarely used example thereof, because of the prohibitively large computational cost of simulating electrically very large irregular scattering systems~\cite{imani2021smart}. In terms of open-loop forward models, there are the two approaches we already encountered in Sec.~\ref{sec_ChannelEstimation}: neural surrogate forward models~\cite{stylianopoulos2022deep,momeni2023backpropagation} and physics-compliant closed-form forward models~\cite{sol2023experimentally}. 

Based on the chosen forward mapping, different types of algorithms may be employed to optimize the RIS configuration. The simplest algorithm is an iterative optimization. This approach is naturally compatible with the constraint from Eq.~(\ref{eqB}). For instance, starting with a random RIS configuration, one can test element after element if switching its configuration reduces $\mathcal{C}$ and keep the change of configuration in that case. Multiple loops over all RIS elements are generally necessary due to the non-linearity of the mapping from RIS configuration to channel. Such kinds of iterative optimizations based on experimental measurements (or sometimes also on numerical full-wave simulations) were the standard approach in the field of RIS-parametrized rich-scattering systems until very recently. Now, these iterative optimizations can be implemented based on neural surrogate forward models~\cite{stylianopoulos2022deep} or physics-compliant closed-form forward models~\cite{PhysFad,hugo_eucap2024,mursia2023modeling,tapie2023systematic}, avoiding the prohibitive cost associated with the closed-loop approaches. 

A simple alternative to iterative optimizations, or an approach to initialize those, is a dictionary search. In its simplest form, it consists in generating a very large dictionary of random RIS configurations and corresponding channels, and then selecting the dictionary entry with the lowest value of $\mathcal{C}$. Given the burden of closed-loop forward mappings, this approach is more attractive when very large dictionaries can be rapidly generated with open-loop forward models~\cite{sol2023experimentally}. For the inverse-design of static nanophotonic metasurfaces, this approach has been reported to be appealing~\cite{nadell2019deep}.

A third approach is the so-called ``adjoint method'' that considers the RIS configuration as trainable weights and backpropagates $\mathcal{C}$ in order to identify a RIS configuration that minimizes $\mathcal{C}$. This approach is well-established for the inverse-design of static nanophotonic structures; the training of ANNs via backpropagation is mathematically closely related to this adjoint method~\cite{peurifoy2018nanophotonic}. The adjoint method requires a \textit{differentiable} forward model, i.e., it is incompatible with closed-loop forward mappings via experiment or full-wave simulation. Moreover, it would generally require continuously tunable polarizabilities for the RIS elements such that it is not straightforwardly compatible with the constraint from Eq.~(\ref{eqB}). Nonetheless, this hurdle can be overcome via tricks such as a ``temperature parameter''~\cite{TemperatureParameter} that gradually tunes the distribution of the available polarizability values from continuous to discrete over the course of the optimization~\cite{del2020learned,saigre2022intelligent}. This approach has not been applied to RIS inside rich-scattering environments to date, but it has successfully been applied to the end-to-end optimization of 1-bit programmable DMAs for task-specific sensing~\cite{del2020learned}. DMAs are another example of a massively programmable complex scattering system that is conceptually very closely related to the RIS-parametrized rich-scattering radio environment.

Besides these three families of approaches based on forward mappings of some kind, there are also approaches based on inverse mappings. We do not discuss those in detail here. As stated above, significant efforts are necessary to address ``existence'' and ``uniqueness'' concerns in these cases, which might be achievable with auto-encoder-like algorithms (coupling inverse and forward mappings), generative adversarial networks or tandem networks. This is uncharted territory for the optimization of RIS under rich-scattering conditions to date. However, a substantial literature on such approaches for the inverse-design of static nanophotonic metasurfaces exists~\cite{wiecha2021deep,khatib2021deep}.

Finally, we note that these broad families of approaches are not mutually exclusive and hybrid approaches combining, for instance, a dictionary search with an iterative optimization, may be appealing.

\subsection{Efficiently Updating Physics-Compliant Channels}
\label{subsec_prod}

We have argued in Sec.~\ref{subsec_tax_opt_strateg} that solving the inverse problem based on open-loop physics-compliant forward models is appealing. Most algorithms (gradient descent, dictionary search, etc.) require many forward evaluations, i.e., evaluations of $\mathbf{H}$ for different $\mathbf{c}$. Based on Eq.~(\ref{eq_27}), one may fear that each forward evaluation implies a significant computational cost since it requires the inversion of an $N_\mathrm{P} \times N_\mathrm{P}$ matrix associated with an algorithmic complexity of $\mathcal{O}(N_\mathrm{P}^3)$. Fortunately, however, there is an efficient way to update a previously evaluated channel matrix without performing another full matrix inversion. Indeed, different channel realizations only differ regarding parts of the diagonal of $\mathbf{W}$ and this insight enables efficient channel updates~\cite{prod2023efficient}.

We assume that we have previously evaluated the channel matrix $\dot{\mathbf{H}}$ corresponding to the RIS configuration $\dot{\mathbf{c}}$. Now, we seek an update of the channel matrix such that it corresponds to a new RIS configuration $\ddot{\mathbf{c}}$ instead. $\dot{\mathbf{c}}$ and $\ddot{\mathbf{c}}$ differ regarding the configuration of $d \leq N_\mathrm{S}$ RIS elements. We denote by $\mathcal{M}$ the set of indices of RIS elements whose configurations differ between $\dot{\mathbf{c}}$ and $\ddot{\mathbf{c}}$, and the $k$th entry of $\mathcal{M}$ is $n_k$. 
The corresponding interaction matrices $\dot{\mathbf{W}}$ and $\ddot{\mathbf{W}}$ differ regarding $d$ diagonal entries. Their difference $\Delta\mathbf{W} = \ddot{\mathbf{W}}-\dot{\mathbf{W}}$ can be expressed as $\mathbf{\Delta W}=\mathbf{UCV}$, where $\mathbf{C} \in \mathbb{C}^{d\times d}$ is a diagonal matrix that contains the $d$ non-zero changes of the inverse polarizabilities and the matrices $\mathbf{U} = \mathbf{V}^T \in \mathbb{B}^{N_\mathrm{P}\times d}$ act as ``selectors'' of the entries of $\mathcal{M}$:
\begin{subequations}
    \begin{equation}
        \left[\mathbf{C}\right]_{k,k'}=\delta_{k,k'}\Delta\alpha_{n_k}^{-1}.
    \end{equation}
    \begin{equation}
        \left[\mathbf{U}\right]_{i,k} = \left[\mathbf{V}\right]_{k,i} = \delta_{i,n_k}.
    \end{equation}
\end{subequations}
Applying the Woodbury identity~\cite{hager1989updating} straightforwardly yields:
\begin{equation}
        \ddot{\mathbf{W}}^{-1} = \left(\dot{\mathbf{W}}+\mathbf{UCV}\right)^{-1} =\dot{\mathbf{W}}^{-1}-\dot{\mathbf{W}}^{-1}\mathbf{U}\left(\mathbf{C}^{-1}+\mathbf{V}\dot{\mathbf{W}}^{-1}\mathbf{U}\right)^{-1}\mathbf{V}\dot{\mathbf{W}}^{-1}.
\label{eq:woodbury}
\end{equation}
However, since we are only interested in $\ddot{\mathbf{H}} \propto \left[\ddot{\mathbf{W}}^{-1} \right]_\mathcal{RT}$, the most efficient approach to obtain $\ddot{\mathbf{H}}$ is~\cite{prod2023efficient}
\begin{equation}
\ddot{\mathbf{H}} \propto \left[\ddot{\mathbf{W}}^{-1}\right]_{\mathcal{RT}} =  \left[\dot{\mathbf{W}}^{-1}\right]_{\mathcal{RT}} - \left[\dot{\mathbf{W}}^{-1}\right]_{\mathcal{RM}}\left(\mathbf{C}^{-1}\!+\!\left[\dot{\mathbf{W}}^{-1}\right]_\mathcal{MM}\right)^{-1}\left[\dot{\mathbf{W}}^{-1}\right]_{\mathcal{MT}}.
\label{eq:reducedwoodbury}    
\end{equation}
The algorithmic complexity of Eq.~(\ref{eq:reducedwoodbury}) is $\mathcal{O}\left(d^3\right)$ for the inner matrix inversion, $\mathcal{O}\left(N_\mathrm{R}d^2\right)$ or $\mathcal{O}\left(N_\mathrm{T}d^2\right)$ for the first matrix product (depending on whether the leftmost or the rightmost is computed first), and $\mathcal{O}\left(N_\mathrm{R}dN_\mathrm{T}\right)$ for the remaining one. 

For the special case of a scattering environment composed of discrete dipoles surrounded by free space that we considered in Sec.~\ref{subsec_DipoleScattEnv}, efficient methods to update the channel matrix upon displacements of wireless entities or changes of the properties of the environmental dipoles exist, too~\cite{prod2023efficient}. Moreover, the Woodbury identity can similarly be applied to RIS configuration updates in the impedance-based physics-compliant model formulations~\cite{tapie2023systematic}.

\begin{important}{Key Take-Home Messages of Sec.~\ref{sec_Optimization}}
\begin{enumerate}
\item RISs serve either for \textit{channel shaping} or \textit{information encoding} in applications spanning from wireless communications via wave-based computing to sensing.
\item Optimizing the RIS configuration (under rich-scattering conditions or not) is an inverse-design problem that can typically not be solved purely based on an inverse model (unlike a sensing problem).
\item RIS optimization under rich-scattering conditions so far largely relied on experimental or full-wave forward mappings of closed-loop nature, combined with iterative optimizations.
\item Recent open-loop forward models, notably the compact closed-form physics-compliant models seen in Sec.~\ref{sec_ChannelModeling} and Sec.~\ref{sec_ChannelEstimation} (but also neural surrogate models), can be efficiently used in iterative optimizations, but also for dictionary searches or adjoint methods.
\item Most optimization algorithms require the evaluation of many forward mappings. In a physics-compliant model, these can be efficiently computed using the Woodbury identity as opposed to inverting the interaction matrix from scratch for each forward mapping.
\end{enumerate}
\end{important}

\section{Summary and Future Research Opportunities}
\label{sec_summary}

In this chapter, we have \textbf{formulated} physics-compliant end-to-end channel models for RIS-parametrized rich-scattering radio environments in Sec.~\ref{sec_ChannelModeling}, we \textbf{estimated their parameters} in unknown complex radio environments in Sec.~\ref{sec_ChannelEstimation}, and we discussed how they can be \textbf{used for open-loop optimization} of the RIS configuration for a desired wireless functionality in Sec.~\ref{sec_Optimization}.

Looking forward, we expect that, among others, the following open research questions will be addressed in the area of RIS-parametrized rich-scattering radio environments:

\begin{itemize}
    \item In terms of the formulations of physics-compliant models, it is important to harmonize existing polarizability-based and impedance-based approaches by formalizing their equivalence via rigorous derivations from first physical principles~\cite{pdh_eucap2024}.

\item The insights derived from the physics-compliant models also raise the question of whether future RIS design efforts should be dedicated to mitigating coupling (e.g., inspired by existing approaches for patch antenna arrays~\cite{wu2017array,li2018isolation,lin2020weak,zhang2021simple,zhang2021novel}) or rather to purposefully engineering it (as suggested by recent ``beyond-diagonal RIS'' (BD-RIS) ideas~\cite{shen2021modeling,li2022beyond,nerini2023discrete}). While this question is not specific to the rich-scattering setting, the latter can be interpreted as a non-tunable randomly connected BD-RIS. On the one hand, it has been observed on various occasions that reverberation under rich-scattering conditions boosts the control of the RIS over the channel~\cite{sol2022meta}; on the other hand, theoretical works highlight similar benefits of (tunable) BD-RIS over a ``diagonal''-RIS in free space. This raises the question whether a randomly connected rather than tunable BD-RIS may constitute a good trade-off between the achievable performance improvement with a BD-RIS and the complexity of the hardware implementation.

\item In terms of estimating the model parameters, it is important to extend the existing single-frequency approach in efficient manners to wideband scenarios, and to develop a theoretical understanding of intriguing features like the phase transition in the dependence of the model accuracy on the number of calibration examples.

\item In terms of optimizing the RIS configuration, the recently unlocked potential of open-loop control with compact physics-compliant models remains largely unexplored.

\item All discussions in this chapter were dedicated to \textit{static} rich-scattering radio environments. The more realistic case in which some of the \textit{wireless entities dynamically move} and possibly change their shapes then implies a non-linear double-parametrization of the wireless channels via the controllable RIS configuration and the uncontrollable motion~\cite{ChloeMag}. These two effects cannot be treated independently from each other, requiring modifications of the model formulation, the channel estimation and the RIS optimization. In particular, we expect that this non-linear double-parametrization constitutes a qualitatively new motivation for integrated sensing and communications (ISAC) in RIS-parametrized dynamic rich-scattering environments~\cite{ChloeMag,zhao2022intelligent}.

\end{itemize}

\bibliographystyle{spphys}

\end{document}